# Implications of the virus-encoded miRNA and host miRNA in the pathogenicity of SARS-CoV-2

Zhi Liu[1,2], Jianwei Wang[1,2], Yuyu Xu[1,2], Mengchen Guo[1,2], Kai Mi[1,2], Rui Xu[1,2], Yang Pei[1,2], Qiangkun Zhang[1,2], Xiaoting Luan[1,2], Zhibin Hu[3], Xingyin Liu[1,2#]

[1]State Key Laboratory of Reproductive Medicine, Department of Pathogen Biology-Microbiology Division, Nanjing Medical University, Nanjing, 211166, China.
[2]Key Laboratory of Pathogen of Jiangsu Province and Key Laboratory of Human Functional Genomics of Jiangsu Province, Nanjing Medical University, Nanjing, 211166, China.
[3]State Key Laboratory of Reproductive Medicine, School of Public Health, Nanjing Medical University, Nanjing, 211166, China.

**Correspondence: Dr. Xingyin Liu, xingyinliu@njmu.edu.cn,**

## Abstract

The outbreak of COVID-19 caused by SARS-CoV-2 has rapidly spread worldwide and has caused over 1,400,000 infections and 80,000 deaths. There are currently no drugs or vaccines with proven efficacy for its prevention and little knowledge was known about the pathogenicity mechanism of SARS-CoV-2 infection. Previous studies showed both virus and host-derived MicroRNAs (miRNAs) played crucial roles in the pathology of virus infection. In this study, we use computational approaches to scan the SARS-CoV-2 genome for putative miRNAs and predict the virus miRNA targets on virus and human genome as well as the host miRNAs targets on virus genome. Furthermore, we explore miRNAs involved dysregulation caused by the virus infection. Our results implicated that the immune response and cytoskeleton organization are two of the most notable biological processes regulated by the infection-modulated miRNAs. Impressively, we found hsa-miR-4661-3p was predicted to target the S gene of SARS-CoV-2, and a virus-encoded miRNA MR147-3p could enhance the expression of TMPRSS2 with the function of strengthening SARS-CoV-2 infection in the gut. The study may provide important clues for the mechisms of pathogenesis of SARS-CoV-2.

## Introduction

The current pandemic of COVID-19, caused by a novel virus strain, SARS-CoV-2, has led to over 1,400,000 confirmed cases and 80,000 fatalities in over 100 countries since its emergence in late 2019. SARS-CoV-2 is an enveloped, positive-sense, single-stranded RNA betacoronavirus of the family *Coronaviridae*. Humans coronaviruses cause respiratory tract infections that can be mild, such as some cases of the common cold, e.g. HCoV-NL63, HCoV-OC43, HKU1. However, over the past two decades, highly pathogenic human coronaviruses have emerged, including SARS-CoV in 2002 and 2003, which infected 8,000 people worldwide and with a fatality rate of ~10% and MERS-CoV in 2012 with 2,500 confirmed cases and a death rate of 37%[1]. Infection with these highly pathogenic coronaviruses can result in Acute Respiratory Distress Syndrome (ARDS), which may lead to widespread inflammation in the lungs, shortness of breath, and death. Compared to MERS and SARS, SARS-CoV-2 appears to spread more efficiently although with relatively lower mortality of 2%~3%[1,2]. Furthermore, mounting evidence has suggested that the subclinical cases, which show limited to no symptoms but could spread virus as well, may represent a large amount of the infections[3,4]. The existence of the large pool of covert cases is a new threat to the control of the SARS-CoV-2 outbreak.

The SARS-CoV-2 entries cells by engaging the angiotensin-converting enzyme 2 (ACE2) as receptor with the surface unit (S1) of the spike (S) protein, and then uses the host serine protease TMPRSS2 for S priming, allowing fusion of viral and cellular membranes and viral entry into the cell[5]. Pneumonia is the common symptom of SARS-CoV-2 infection[6]; however, there was also evidence of damages to the liver[7] and spleen[8] as well as gastrointestinal symptoms[9]. The SARS-CoV-2 infection causes increased secretion of cytokines, and the main death cause of SARS-CoV-2 is ARDS, for which one of the main mechanisms are the cytokine storm, the deadly uncontrolled systemic inflammatory response resulting from the release of large amounts of pro-inflammatory cytokines and chemokines by immune cells in SARS-CoV-2 infection[6]. Nevertheless, the mechanism of virus invasion, release and the causes of these exuberant inflammatory responses in SARS-CoV-2 infection remain largely unknown.

miRNA is a small non-coding RNA molecule with an average of 22 nucleotides in length. miRNA was widely found in plants, animals and some viruses and involved in a variety of biological processes. In most cases, miRNAs interact with the 3′ untranslated region (3′ UTR) of target mRNAs to induce mRNA degradation and translational repression. However, except for the role as a repressor, increasing evidence has shown that miRNAs could activate gene expression by interacting with other regions, including the 5′ UTR and gene promoters[10,11]. As crucial regulators of gene expression, miRNA-targeted therapeutics has been proposed or for treatment of cancers, virus infection and other diseases. For example, antimiR targeted at has-miR-122 is in trials for treating hepatitis[12].

It has shown that viral genomes, including DNA and RNA virus, were capable of encoding miRNAs[13,14]. The virus-derived miRNAs can be expressed in host cells and participate in the lifecycle and cellular consequences of infection. In 2004, Tuschl and colleagues identified the first viral miRNAs encoded by Epstein-Barr virus (EBV), a common human herpesvirus types in the herpes family. To date, over 300 viral encoded miRNAs have been described. A couple of EBV-encoded miRNAs were identified since 2004 and were proposed to be involved in the host immune response[15]. More than 20 human cytomegalovirus (HCMV) miRNAs are identified and suggested to target several cellular genes to evade the immune system, control cell cycle and vesicle trafficking[16,17]. The HIV encodes miRNAs, i.e. MiR-H3 and HIV1-miR-H1 were reported to enhance viral production and impairs cellular responses to infection, respectively[18,19].

Viral miRNAs have been found to target a large number of host genes involved in regulating cell proliferation, apoptosis, and host immunity[13]. The miR-HA-3p encoded by H5N1 accentuates the production of antiviral cytokines by targeting and repressing poly(rC)-binding protein 2 (PCBP2) expression, a known negative regulator of RIG/MAVS signaling, resulting in a high level of cytokine production and high mortality[20]. Also, viral miRNAs target the virus genes, which also help the virus to remain hidden from the host immune response. For example, miR-BART2 encoded by EBV was found to lie antisense to the EBV DNA polymerase gene BALF5 and was proposed to inhibit DNA polymerase expression by inducing cleavage within the 3'-UTR of BALF5[21]. Furthermore, given the importance of miRNA in regulating gene networks or pathways, viruses have evolved mechanisms to degrade, boost, or hijack cellular miRNAs to benefit the viral life cycle. Both Herpesvirus saimiri (HVS) and Murine cytomegalovirus (MCMV) bind miR-27 with antisense, leading miR-27 degradation in a sequence-specific and binding-dependent manner, to promote T-cell activation and enhance viral replication, respectively[22,23]. The non-translated region of North American eastern equine encephalitis virus genome base pairs with the miR-142-3p, resulted in a miR-142-3p mediated innate immune suppression[24]. Moreover, recent studies indicate the virus-encoded miRNAs, host-encoded miRNAs, and miRNA targets together form a novel regulatory system between the virus and the host, which contributes to the outcome of infection[25]. Hence, it is crucial to comprehensively understand the role of the novel regulatory system during SARS-COV-2 infection, then apply this information towards developing both new medicine and repurposing existing ones.

## Materials and Methods

### Computational identification of SARS-CoV-2-encoded miRNA

The complete genome sequence of SARS-CoV-2 (accession number: MN908947) was obtained from the National Center for Biotechnology Information (NCBI). The VMir software package (v2.2)[26], an ab initio prediction program designed specifically to identify pre-miRNAs in viral genomes, was used to identify and visualize potential pre-miRNAs within the SARS-CoV-2 genome. The program examines all possible pre-miRNAs in the input sequence and consequently identifies a large number of potential candidates. The program slides a window of defined size across the entire sequence of interest, advancing each window by a given step size. The parameters of window and step sizes were set at 500 and 1 nt, respectively, as set in other literature. The Minimum Free Energy of the predicted hairpin structures were predicted with RNAFold algorithm from the Vienna package[27]. Hairpins with MEF > -20 kCal/mol were removed from the flowing analysis. Then HuntMi software was used to further distinguish miRNA hairpins from pseudo hairpins with a virus-based model[28]. Maturebayes web server[29] was used to identify the mature miRNA within the pre-miRNAs based on sequence and secondary structure information of their miRNA precursors.

### Prediction of miRNA targets and functional annotation

Human 3′- and 5′-UTR, and the Enhancer sequence for five tissues were downloaded from the UTRdb database[30] and EnhancerAtlas(35) database, respectively. Human miRNA sequence was downloaded from the TargetScan database. The targets of virus-encoded miRNA and human miRNAs were predicted with TargetFinder[31] and miRanda[32]. The miRNA targets on human genes were extracted from the TargetScanHuman 7.2 database[33], with context++ score < -0.4 and percentile >0.99 as previously applied[34]. Then UTR and enhancer associated genes were extracted from annotations in UTRdb database[30] and EnhancerAtlas database[35]. Genes were functionally annotated using Gene Ontology (GO) terms using R packages and the function enrichment tests were conducted with GOstats[36].

### Virus variation processing

The genomic variations were downloaded from China National Center for Bioinformation, data were collected until Mar 19, 2020. The mutations were obtained based on sequence alignment against the reference genome MN908947.3 as we used for miRNA scanning.

### Data process for other coronaviruses

The genome sequence of SARS-CoV, MERS-CoV-2, and HCoV-NL63 were obtained from NCBI with the accession of NC_004718.3, NC_005831.2, and NC_019843.3, respectively. The pipeline for miRNA prediction from the virus genome, miRNA target prediction, and functional annotation was the same as that applied to the SARS-CoV-2 as we described before.

## Results and Discussion

### Identification of SARS-CoV-2-encoded miRNAs

A total of 808 potential pre-miRNA in the SARS-CoV-2 genome were detected with VMir Analyzer[26], an ab initio prediction program designed specifically to identify pre-miRNAs in viral genomes (Figure 1A). Candidate pre-miRNA sequences were further identified using HuntMi[28] with the virus model and the sequences with minimum free energy (MFE) > -20 kCal/mol were removed. A total of 45 virus pre-miRNA candidates were finally obtained, with 30 were found in the direct orientation and 15 in the reverse orientation. The pre-miRNAs were evenly distributed across the SARS-CoV-2 genome (Figure 1B). The average length of the pre-miRNA sequence was 78 nt with an average MFE of -28.1 kCal/mol. Finally, the sequence and position of mature miRNA within the pre-miRNA candidate were identified with MatureBayes, resulting in 90 mature miRNAs.

### Virus miRNAs suppress host gene expression

The classical mechanism of miRNAs to regulate their target gene is by binding to the 3′ UTR of mRNA to exert negative regulatory effects on gene expression. Human 3′ UTR targets of the mature viral miRNAs were predicted and 40 miRNA were predicted to be binding to the 3′ UTR of 73 genes. Gene ontology analysis revealed that the top functions of the target genes were enriched in the function of Notch binding, single-stranded DNA endodeoxyribonuclease activity, deoxyribonuclease activity, cellular response to peptide hormone stimulus and regulation of fatty acid metabolic process (Figure 2A). One of the common related biological processes involved by the notch signaling pathway, DNA endodeoxyribonuclease, and deoxyribonuclease is apoptosis, which is considered as a powerful mechanism to curtail viral spread[37]. However, consequently, viruses have evolved sophisticated molecular strategies to subvert host cell apoptotic defenses. A couple of studies have investigated the mechanism by which viruses modulating apoptosis signaling[16,38]. The targeted gene annotated to the related GO items included CHAC1 and RAD9A among others, which were targeted by the virus-encoded miRNA, namely MD2-5p and MR147-3p, respectively (Figure 2B). CHAC1 (cation transport regulator-like protein 1) is a proapoptotic enzyme and a regulator of Notch signaling. RAD9A is a cell cycle checkpoint protein required for cell cycle arrest and DNA damage repair in response to DNA damage. RAD9A interacts with BCL-2/BCL-xL, the most important genes that regulate cell death, and promote apoptosis[39]. The suppressive role of the SARS-CoV-2-encode miRNAs on these genes suggested the possible role of them in reducing the host cell apoptotic to subvert host defense.

On the other hand, the involvement of the targeted genes in the biological process of *cellular response to peptide hormone stimulus* and *regulation of fatty acid metabolic process* might imply the role of them in the pathological damage. Genes enriched in these biological processes included FOXO3, ADIPOQ, and ADIPOR1, among others. The FOXO family represents a group of transcription factors that are required for many stress-related transcriptional programs including antioxidant response, gluconeogenesis, cell cycle control, apoptosis, and autophagy[40]. FOXO3 is required for antioxidant responses and autophagy, and altered expression of FOXO3 was observed in Hepatitis C infection and fatty liver[41]. ADIPOQ is a gene encodes adiponectin, which regulates both glucose and lipid metabolism and exerts an insulin-sensitizing effect in the liver. The binding of adiponectin with its specific receptors, i.e. ADIPOR1/2, induces the activation of a proper signaling cascade that becomes altered in liver pathologies[42]. Zhang et. al reported that 2–11% of patients with COVID-19 had liver comorbidities and 14–53% cases reported abnormal levels of alanine aminotransferase and aspartate aminotransferase (AST) during disease progression[7]. However, it is unclear whether the liver damage of COVID-2019 patients is caused directly by the viral infection or by the drug toxicity. The analysis implied the liver dysfunction may be impaired by SARS-COV-2 virus infection through the release of virus miRNA.

## Virus miRNA activate host gene expression

Besides the repression role of miRNA on the target genes, miRNAs were also found to target at promoter regions of genes to activate their expression[43]. Therefore, we also searched the targets of SARS-CoV-2 encoded miRNA on the 5′ UTR of human genes. The 11 virus miRNAs were identified to be bind to the 5'-UTR of 13 target genes, including the binding between MR385-3p and TGFBR3 (Transforming Growth Factor Beta Receptor 3) is a gene widely expressed on cells of both the innate and adaptive immune system and it is reported to play a role in promoting Th1 differentiation and regulation of regulatory T-cell activation and survival[44].

Accumulating evidence has proved that miRNA can also activate gene expression by targeting enhancers region in the nucleus[11]. We then downloaded the tissue-specific enhancer sequences from the EnhancerAlta for five tissues which were reported to be affected by SARS-CoV-2 infection, i.e. lung, gut, spleen, liver, and heart. Target prediction revealed that the lung is with the largest number of genes targeted by miRNA on the enhancer, followed by spleen and gut (Figure 3A).

Functional annotation of these targeted genes revealed a notable enrichment in chemokine signaling pathways and cell skeleton related functions, such as actin filament-based process, actin cytoskeleton organization and cytoskeleton organization in the lung (Figure 3B). For example, MR147-5p targeted at the putative enhancer region of CXCL16 and ARRB2. CXCL16 was reported to regulate cell-mediated immunity to salmonella enterica serovar snteritidis via the promotion of gamma Interferon (IFN-γ) production[45]. ARRB2 (arrestin beta-2) plays an important role in inflammation, the deficiency of it abolishes ovalbumin-evoked T lymphocyte and eosinophil infiltration of the lungs and eliminates Th2 cytokine production in the lungs. Another category of genes was related to the cell cytoskeleton organization, e.g. MYH9 (myosin-9) and ITGB5 (Integrin beta-5). Actin cytoskeleton is critical for viral replication at many stages of the viral life cycle, and viruses can subvert the force-generating and macromolecular scaffolding properties of the actin cytoskeleton to propel viral surfing, internalization, and migration within the cell[46,47]. These observations suggested the possible role of SARS-CoV-2 encoded miRNA in causing increased inflammation in the lung and facilitating the virus invasion.

In the spleen, the target genes were mainly related to the inflammatory response, i.e. regulation of chronic inflammatory response and positive regulation of chronic inflammatory response to antigenic stimulus (Figure 3C). For example, MR66-3p was predicted to bind to the enhancer of TNF-α, one of the most important cytokines in the "cytokine storm".

In the gut, 54 genes were predicted to be enhanced by 34 miRNAs. The most notable target of the virus miRNA is TMPRSS2, which is reported to enhance SARS-CoV-2 infection together with ACE2[48]. The enhancer of TMPRSS2 is targeted by MR147-3p in the gut. Several studies have implicated the gastrointestinal infection of SARS-CoV-2. Early reports showed that 2–10% of patients with COVID-19 had gastrointestinal symptoms such as diarrhea, abdominal pain, and vomiting[9,49]. SARS-CoV-2 RNA has also been detected in the stool of patients[50,51]. The putative activation role of virus-derived MR147-3p might add evidence to the gastrointestinal infection of SARS-CoV-2 and provided possible mechanisms for the effective evasion of SARS-CoV-2 into gut cells. The most significant enriched gene function is "transport" (Figure 3D). Transporters involved in the movement of ions, small molecules, and macromolecules, such as another protein, across a biological membrane, and play an important role in the maintenance of intestine equilibrium[52].

In the liver, the virus miRNA mainly regulates genes involved in the function of actin filament severing and regulation of cellular protein metabolic process (Figure 3E). MR198-3p act on the enhancer of ADAR, an adenosine deaminases that act on RNA. ADAS is also indicated to act as a suppressor of IFN system responses in various virus infections[53]. It has been reported that a substantial proportion of COVID-19 patients showed signs of various degrees of liver damage, the mechanism and implication of which have not yet been determined. The multiple targets of SARS-CoV-2 encoded miRNA on the enhancer of genes which were highly expressed in the liver implicated a possible role of virus miRNA on the liver damage of COVID-19.

The MYH9 (non-muscle myosin heavy chain 9) was enhanced by SARS-CoV-2 derived MR359-5p in all the detected tissues except for the heart. In a study on PRRS virus (Porcine reproductive and respiratory syndrome virus), a positive sense, single-stranded RNA virus that causes highly infectious porcine disease, MYH9 has been proved to provide a bridge between the attachment of virus particles to cell surface receptors and the subsequent un-coating events required for genomic release within the host cell[54]. RXRA is another gene that may be activated by virus-encoded miRNA MR328-5p in the four tissue mentioned above. A study has shown that RXRA over-expression or ligand activation increases host susceptibility to viral infections in vitro and in vivo by attenuates host antiviral response through suppressing type I interferon[55]. The putative upregulation of MYH9 and RARA by virus-encoded miRNA might provide additional evidence for the invasion of SARS-CoV-2 to the four human tissues, as well as a mechanism by which the viruses enter and release from the cells and escape from the innate immune system.

## Virus genome hijacks host miRNA to regulate immune response

Several studies have reported that the virus genome can hijack host miRNA to modulate host biological processes. For example, the virus transcript HSUR of Herpesvirus saimiri (HVS) base pairs with the host miRNA 27 (miR-27), leading to miRNA degradation in a sequence-specific and binding-dependent manner, and subsequent activation of infected T cells; HCMV produces a bicistronic mRNA that mediates degradation of cellular miR-17/miR-20a family miRNAs similarly.

A total of 28 human miRNA were predicted to target at the SARS-Cov-2 genome. Since these miRNAs were hijacked by the virus genome, we could suppose that the genes that originally targeted by these miRNA could be affected. There were more than human 800 genes were predicted to be regulated by these miRNA (Figure 4A), and a notable enrichment at the immune system process was observed (Figure 4B). Among the hijacked miRNAs, miR-146 was reported as one of the key modulators of the immune response. Studies in mice have shown that miR-146b-/- mice displayed enlarged spleens, increased myeloid cell populations both in spleen and bone marrow and spontaneously develop intestinal inflammation[56]. Another miRNA, miR-939, though no target was predicted, was proposed to regulate proinflammatory genes by experiment. Increasing miR-939 levels should restore homeostasis by decreasing inflammatory protein synthesis[57]. Our result posed a possibility that the hijack of human miRNA by the SARS-Cov-2 genome might contribute to the abnormal immunity activation in patients with COVID-19.

## Virus genomic region targeted by virus and host miRNA

The SARS-Cov-2 genome consists of 11 open reading frames (ORF), including ORF1ab, which encoding 16 nonstructural proteins, followed by those encode structure proteins, i.e. S (spike protein), E (envelope protein), M (membrane protein), and N (nucleocapsid protein), and six accessory proteins (3a, 6, 7a, 7b, 8, and 10)[58]. Previous studies have shown that miRNAs derived from both virus genome and host can target viral transcript and regulate the virus infection[14]. We then explore the targets of both virus and host cell-derived miRNAs on the SARS-CoV-2 genome.

There were 27 SARS-CoV-2 encoded miRNA that can target the virus genome (Figure 5A). Forty-three targets were predicted for the 27 miRNAs and most of them were bind to the ORFlab region (Figure 5B), the longest ORF in the SARS-Cov-2 genome. There were 2 miRNA targets at the 5′ UTR of the virus genome, and 2 at the S gene, which encodes a spike glycoprotein to bind its receptor ACE2 on human cells, and mediates membrane fusion and virus entry(5).

Twenty-eight human miRNAs were predicted to have 30 targets on the SASRS-CoV-2 genome (Figure 5B). Most of the human miRNAs were bind to the ORF1ab, where the enzymes for virus replication and translation were encoded. Followed by 7 miRNA targeted at the N genes of the virus, and 2 at the 5′ UTR. It has been reported that miR-323, miR-491, and miR-654 inhibit replication of the H1N1 influenza A virus through binding to the polymerase PB1 gene[20]. And miR-122 binding to the HCV genome alters the structure of the 5′ UTR in a manner that promotes viral RNA translation. A human miRNA, hsa-miR-4661-3p, was predicted to target at the genomic region 25,296-25,320 within S genes (21,563-25,384), which is the potential 3′ UTR of the S gene transcript (Figure 5C). This observation suggested a possible repressor role of hsa-miR-4661-3p on the SARS-CoV-2 S gene expression. It might be an example of the antiviral mechanism except for immune response adopted by the host to defense virus.

These results implicated the possible role of both virus and host cell-derived miRNA in the life cycle and pathogenicity of SARS-CoV-2. Since the transcriptome landscape of the SARS-Cov-2 is largely unknown, further researches in characterizing the genomic and transcriptomic features of it may enlighten the role of miRNAs in regulating the virus gene expression and infection.

### Mutations in virus genome affect miRNA function

Mutations accumulated since the SARS-CoV-2 emerged, studies have implicated that the mutations on the virus genome may lead to an attenuated phenotype of SARS-CoV-2. We, therefore, explored whether miRNA encoded by either virus or human contributed to the fitness of SARS-CoV-2 to human.

A total of 810 mutations on the SARS-CoV-2 genome were collected from China National Center for Bioinformation, including 646 SNPs, 19 indels and 145 deletions (Figure 1B). There were 33 pre-miRNAs located within the mutant region, i.e. 30 pre-miRNAs overlapped with SNPs and 4 located in the deletion region, and the number of mutations within each pre-miRNA ranges from 1 to 5. We then examined the effect of genomic mutations on the generation of miRNA in SARS-CoV-2. MFE is an indicator of the hairpin stability of pre-miRNA, the stability of pre-miRNA hairpin structure will be reduced with the increase of MFE. Compared with the wild-type pre-miRNA, the MFE of the mutant pre-miRNAs were significantly increased (paired Wilcox test *p-value* = 3.585e-05, Figure 6A), indicating an overall decreased stability of the mutant-genome encoded pre-miRNA. The most affected pre-miRNA is MR369 with MFE increased from -21.7 kCal/mol to -14.7 kCal/mol, which was considered to not be able to form stable hairpin structure after introducing the mutations (Figure 6B).

We then focused on the miRNAs with increased MFE scores and those located in the deletion regions. In the lung, human genes targeted by these miRNAs at enhancers were enriched in the function of positive regulation of MHC class I biosynthetic process, which refers to two genes, i.e. CITTA and NLRC5. CITTA is targeted by MR288-5p, and NLRC5 by MD202-5p. The protein encoded by CIITA is located in the nucleus and acts as a positive regulator of class II major histocompatibility complex gene transcription, and is referred to as the "master control factor" for the expression of these genes. NLRC5 is a member of the NLR family that acts as a transcriptional activator of MHC class I genes. Silencing of NLRC5 resulted in increased IL-6, TNF, CXCL5, and IL-1β secretion，at the same time decreased secretion of the anti-inflammatory cytokine IL-10[59]. The intact pre-miRNA is predicted to generate mature miRNA that enhancer the gene expression of both CIITA and NLRC5, resulting in an enhanced inflammation. While in the mutant genome, the generation of these miRNAs might be disturbed, thus the enhanced effect was relieved. Additionally, in the spleen, a couple of target genes were enriched in the cytokine-mediated signaling pathway that was missed by the mutant pre-miRNA, including CCR6, OAS1, POMC, XAF1, TNFAIP3, CANX, NLRC5, SQSTM1, and CD27. These results suggested that the mutation in the region overlapped with miRNA might contribute to the attenuated phenotype of SARS-CoV-2 by changing the miRNA repertories encode by the virus.

Next, we further explore the alteration of host miRNA target on the mutation virus genome.

Twenty-two human miRNA can bind to the mutant genome, with 16 overlapped with those binding to the wild type genome (Figure 6C). The genes targeted by the newly recruited miRNA were annotated to the function of epithelial cell differentiation, actin filament fragmentation and so like, while the gene targeted by the miRNA specific binding to the wild type genome were annotated to the immune response-related progress (Figure 6D). The two miRNA as we described before, i.e. hsa-miR-939-5p and hsa-miR-146b-3p, which were involved in maintaining homeostasis by decreasing inflammatory, were predicted to be not capable of binding to the mutant wild type genome. Taken together, we speculate that the mutant type of virus might relieve the high level of immunity response, even the cytokine storm, which leads to the serve symptom or dead of COVID-2019, but on the other hand, enhance the invasion of the virus by intensifying the regulation of the epithelial differential and actin filament fragmentation, through attracting different batches of host miRNAs.

### miRNA view of the difference between SARS-CoV-2 with other human coronaviruses

Comparing with MERS-CoV and SARS-CoV, SARS-CoV-2 appears to have higher transmission rates, but lower mortality[1,2]. To explore whether and how miRNA implicated in this difference between them, we predicted the miRNAs encoded by MERS-CoV, SARS-CoV, as well as HCoV-NL63, a coronavirus that causes mild symptoms in human and uses the same acceptor with SARS-CoV and SARS-CoV-2 for cellular entry (ACE2), using the same pipeline as applied SARS-CoV-2. The largest putative virus-encoded miRNA repository, i.e. 212 mature miRNAs, was detected from the MERS-CoV genome (Table 1), followed by SARS-CoV with 168 miRNAs and SARS-CoV-2 with 90 miRNAs. HCoV-NL63 encodes the least number of miRNAs (64 miRNAs). Furthermore, the number of targets on the virus genome binding by the host miRNA follows the same trend (Table 1).

Functional analysis revealed that high enrichment of genes targeted by the virus-encoded miRNAs in the metabolic process was observed in SARS-CoV and MERS-CoV, but not in SARS-CoV-2 and HCoV-NL630 (Figure S1A). Enrichment in inflammatory and leukocyte was observed in SARS-CoV-2 and MERS-CoV-2, respectively (Figure S1A). For the genes “hijacked” by virus genome, the functional keywords were enriched in chemotaxis, cytokine, chemokine and immune for the entire four viruses studied (Figure S1B), and HCoV-NL63. But the miRNAs targeted at the virus genome were largely different from each other (Figure S2). These results implicated that differences in miRNA repositories and miRNA-target interaction might be involved in the pathological difference among the four coronaviruses.

## Conclusions

In summary, this silico study revealed a systematic view of the possible role of the virus-encode miRNAs as well as the host-derived miRNAs influenced by the SARS-COV-2 infection, which provide insights from the perspective of miRNA into the biological process involved by the infection (Figure 7). Impressively, the systematic analysis revealed that the immune response is the function most affected by the virus-infection introduced miRNA change in repository and targets, which might contribute to the immune evasion of the virus as well as the abnormal immunity activation in the host. The SARS-CoV-2-encoded miRNA is also implicated in the cytoskeleton dynamics that facilitating the virus envision, trafficking within the cell and release. The virus-encode miRNAs were also predicted to be able to repress the expression of genes involved apoptosis and regulating fatty acid metabolic process by binding at the 3′ UTR of host genes. Additionally, the host miRNAs hijacked by the virus genome were mostly involved in the immune system process, such as hsa-miR-146b and hsa-miR-939. Furthermore, the host miRNA hsa-miR-4661-3p was predicted to bind at the potential 3′ UTR of the S gene transcript with the possible role of a repressor on the expression of S gene. Finally, the genomic mutation-introduced alterations on the coding ability of virus miRNA and the targets of both virus and host miRNA might contribute to the attenuated phenotype of SARS-CoV-2 during its evolution. Finally, the comparison of miRNA repository and targets difference between SARS-CoV-2 with other human coronaviruses implicated the role of miRNAs in the shared and distinct clinical characteristics of them. With further experiments to validate the role of candidate miRNAs in vivo, it is promissing to provide deep insight into the mechisms of SARS-CoV-2 pathogenesis from the pespective of miRNA.

**Author contributions**

XL conceived the project. XL and ZL design the project. ZL, XL, JW, YX, MG, and KM performed analysis. All authors did data analyses and interpretations; ZL and XL prepared and finished the manuscript.

**Acknowledgments**

This work was supported by NSFC grant 81671983 and 81871628 to XL, NSFC grant 81703306 to ZL, NSFC grant 81902027 to JW and Young scientist funding from Jiangsu province to JW(BK20171045).

**Competing interests:** Authors declare no competing interests.

**Tables**

Table 1. The number of miRNAs encoded by virus genome and genomic targets bonded by host miRNA for four human coronaviruses.

| Features | SARS-CoV-2 | SARS-CoV | MERS-CoV | HCoV-NL63 |
|---|---|---|---|---|
| **Genome length** | 29903 | 29751 | 30119 | 27553 |
| **Vmir predict** | 808 | 827 | 862 | 740 |
| **Pre-miRNA candidate** | 45 | 84 | 106 | 32 |
| **Mature miRNA** | 90 | 168 | 212 | 64 |
| **Hits by human miRNA** | 30 | 39 | 70 | 21 |

Figure 1

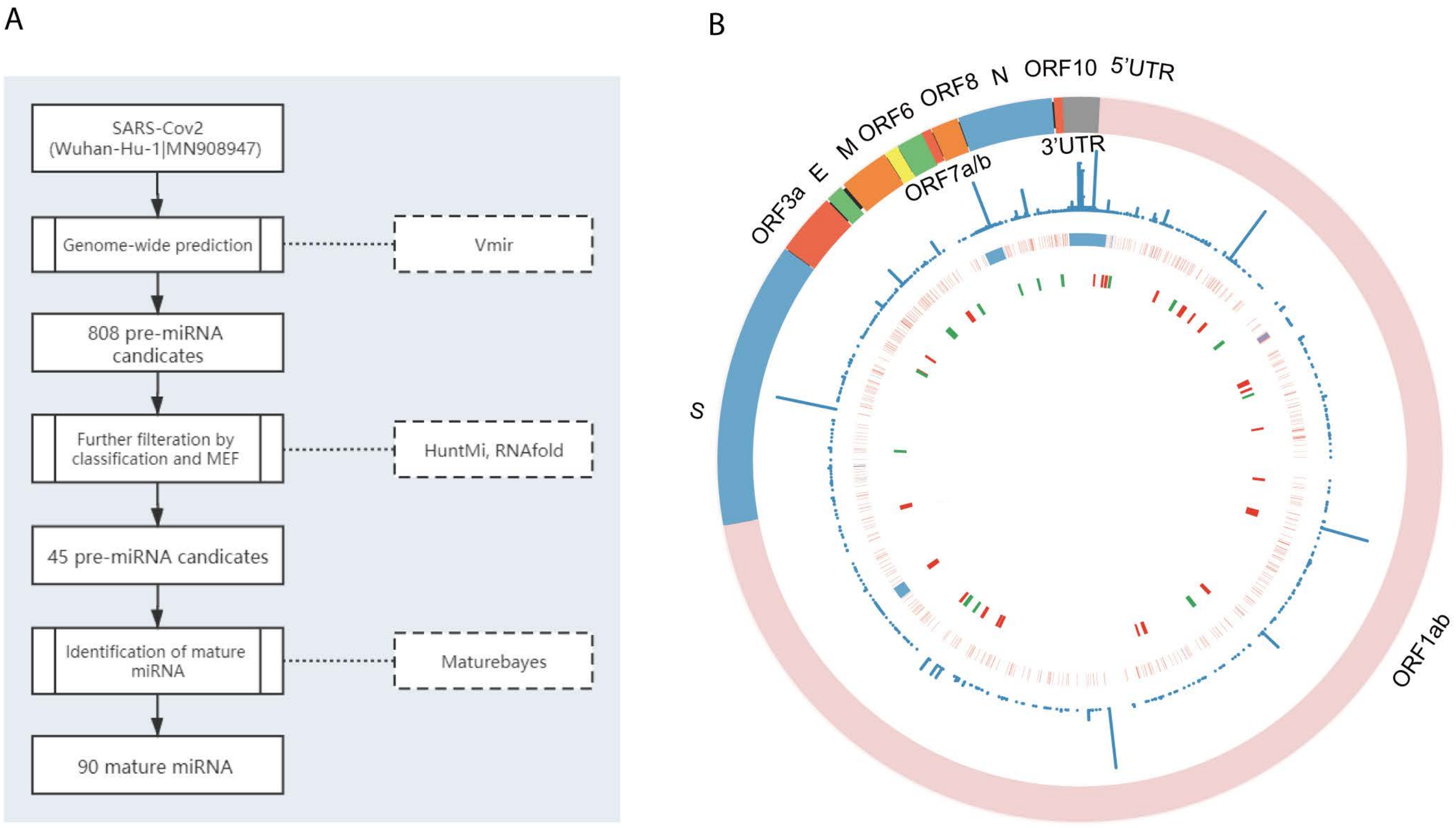

Figure 1. Overview of SARS-CoV-2 encoded miRNA. (A) The schematic workflow for virus miRNA identification. (B). The innermost cycle demonstrated the distribution of pre-miRNA across the SARS-CoV-2 genome with red and green represent pre-miRNAs in the direct direction and reserve direction, respectively. The middle layer demonstrated the mutations in the genome with red denotes SNPs and blue denotes deletions. Outside of it is the frequency of the mutations.

Figure 2

A

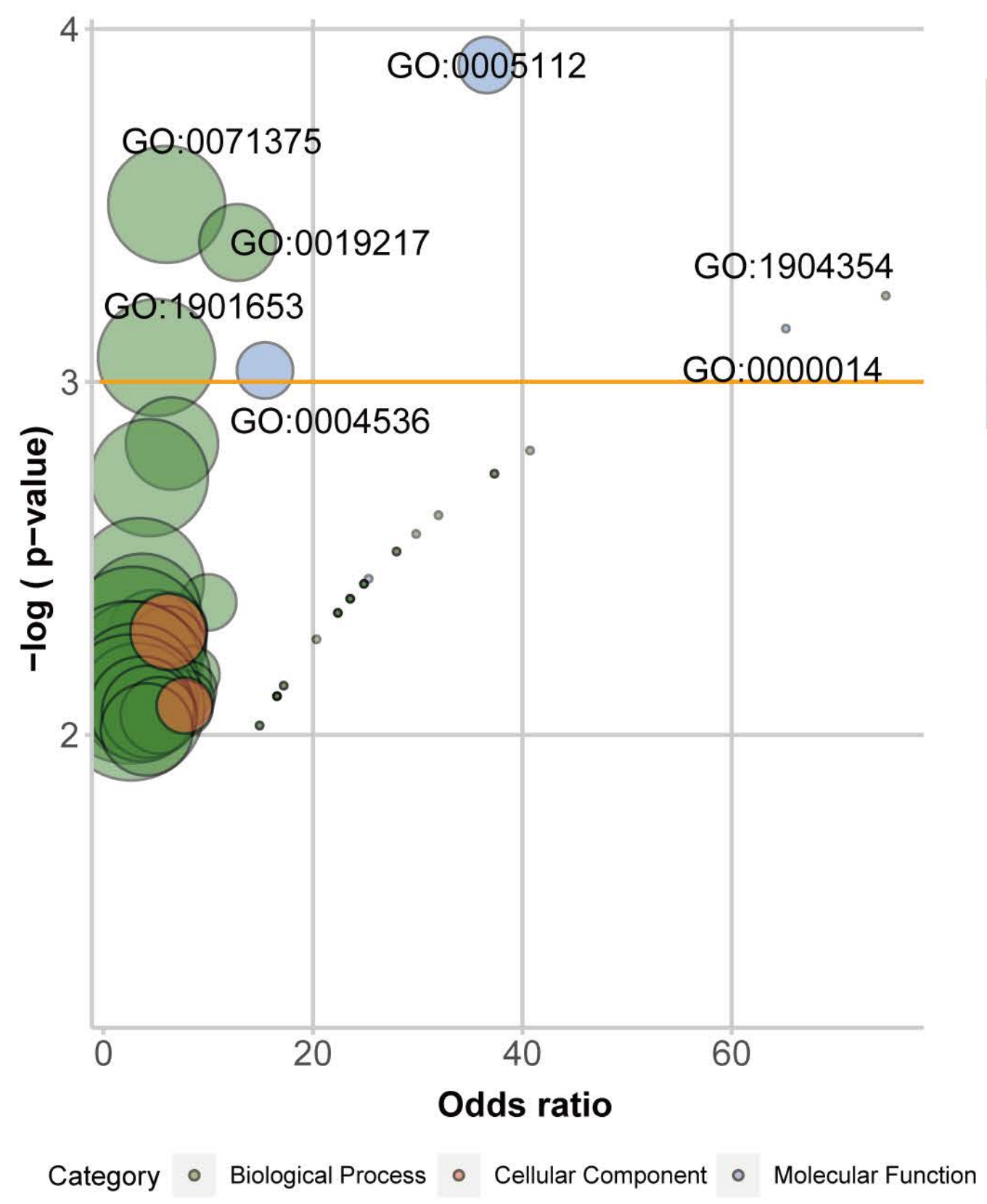

| ID | Description |
|---|---|
| GO:0005112 | Notch binding |
| GO:0000014 | single−stranded DNA endodeoxyribonuclease activity |
| GO:0004536 | deoxyribonuclease activity |
| GO:0071375 | cellular response to peptide hormone stimulus |
| GO:0019217 | regulation of fatty acid metabolic process |
| GO:1904354 | negative regulation of telomere capping |
| GO:1901653 | cellular response to peptide |

B

```
  MD2-5p:    3’  ugGAGGAGGUGCCUCAGAGGUu 5’
                   ||||  |||  |||||||||
  CHAC1:     5’  acCTCC-CCAGGGAGTCTCCAa 3’

  Energy:   -40.450001 kCal/Mol

MD147-3p:    3’  guCCCUGAUGAGGGUGGGUUCu 5’
                   ||||||  |  |||||||||
  RAD9A:     5’  ctGGGACT-CCCCCACCCAAGa 3’

  Energy:   -35.950001 kCal/Mol
```

Figure 2. SARS-CoV-2 encoded miRNA targeting at the 3’ UTR of genes. (A) The GO functions enriched by the virus-miRNA regulated genes. GO IDs with pvalue < -10e-3 were labeled, and the size of babbles is proportion to the number of genes annotated to the corresponding GO terms. (B) Diagram examples of the binding between virus-encoded miRNAs with 3’ UTR.

Figure 3

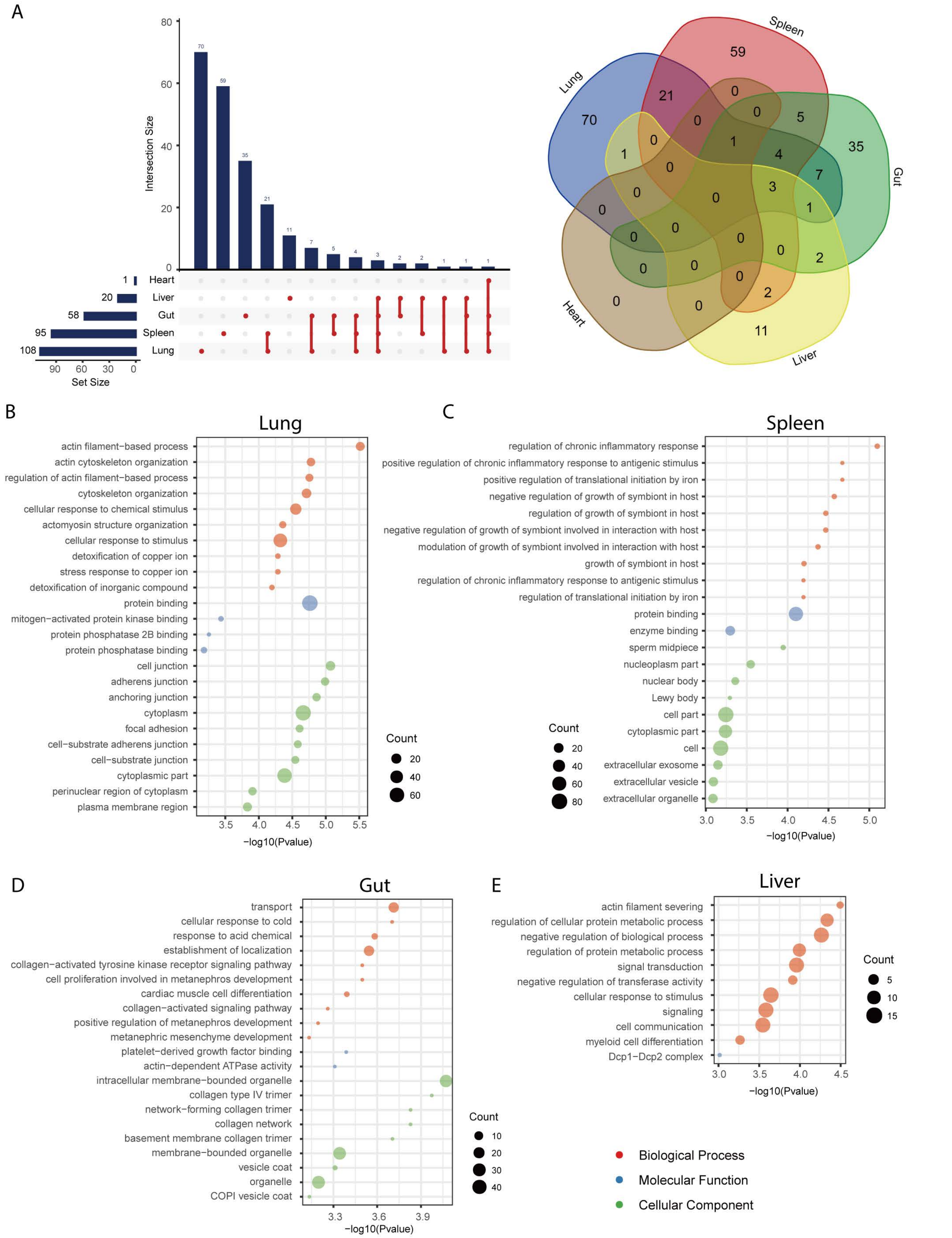

Figure 3. SARS-CoV-2 encoded miRNA targeting at the enhancers in different tissues. (A) The number of miRNA-gene regulation pairs in each tissue and the intersection between different tissues. (B) Functional enrichment genes targeted by the virus miRNA on enhancers in the lung. (C) Functional enrichment genes targeted by the virus miRNA on enhancers in the spleen. (D) Functional enrichment genes targeted by the virus miRNA on enhancers in the gut. (E) Functional enrichment genes targeted by the virus miRNA on enhancers in the liver. GO terms with pvalue < -10e-3 were plotted, and if the number of significant GO terms larger than 10, the top 10 are plotted.

Figure 4

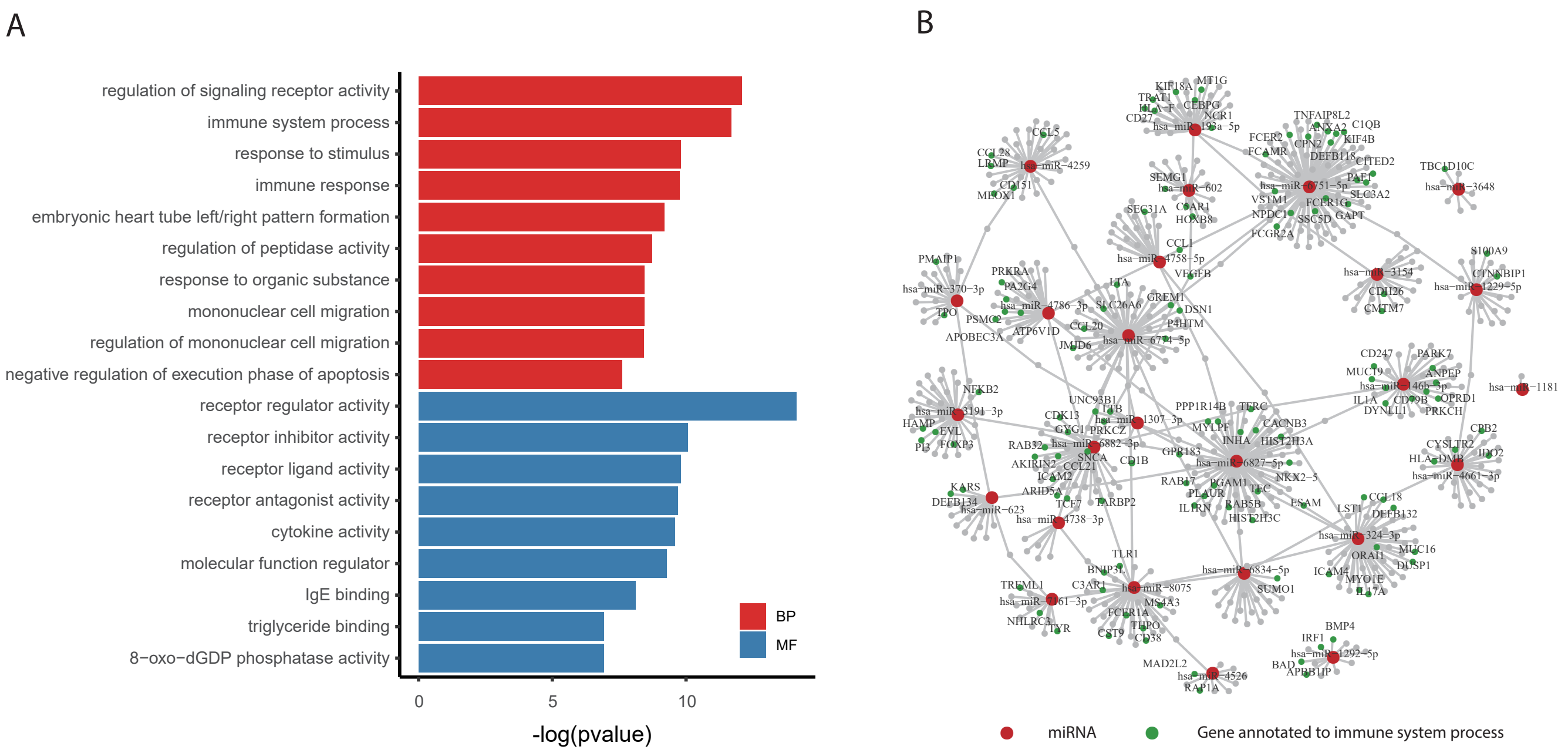

Figure 4. The function of genes targeted by the virus-genome-hijacked miRNAs. (A) The barplot for the enriched functions of genes regulated by the host miRNA attracted by the virus miRNA. (B) The network of the regulation between the hijacked human miRNA and genes. The red nodes represent the miRNA. The greens are genes annotated to the immune system processes, and greys are annotated to other functions.

Figure 5

A

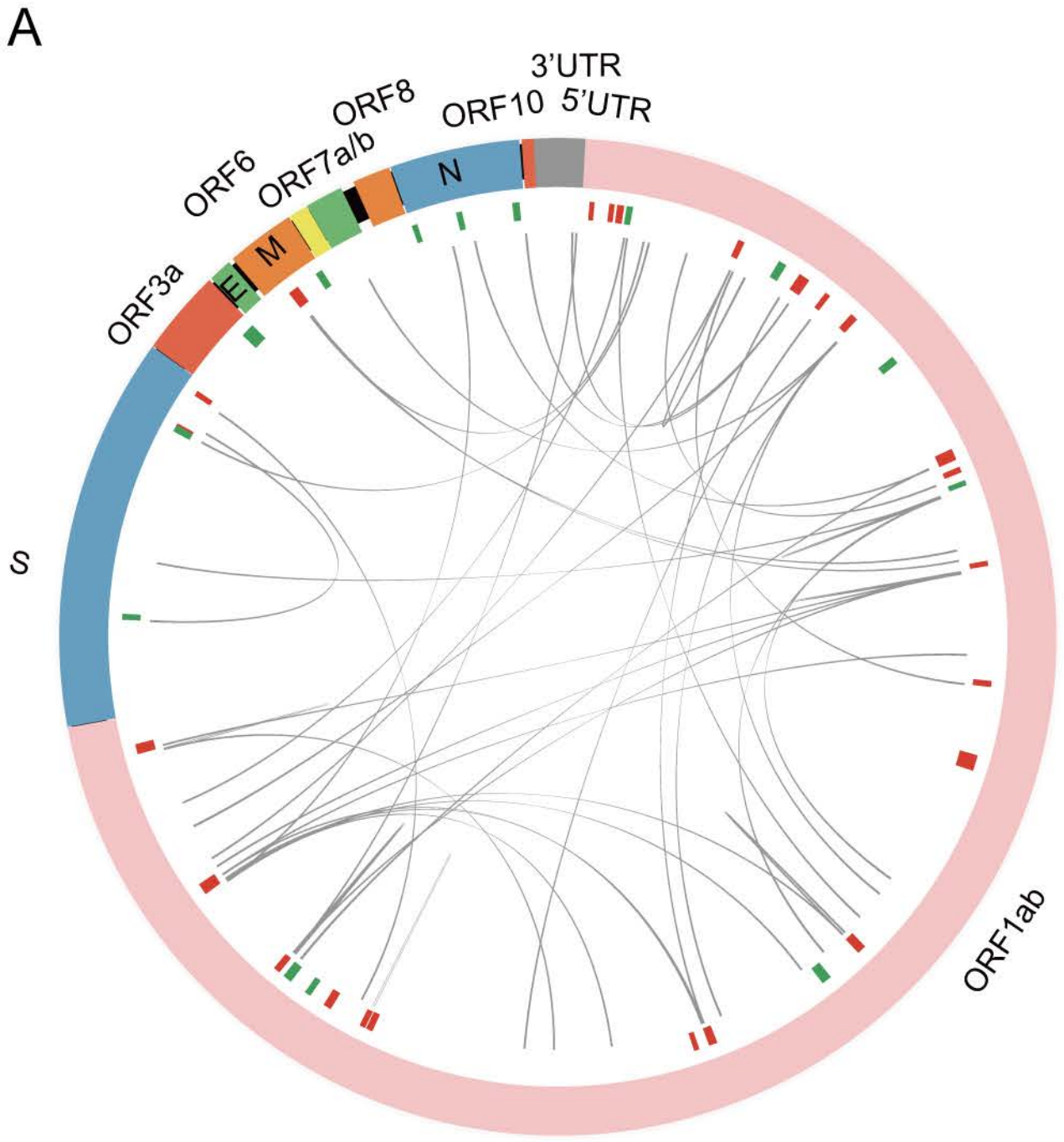

B

| Genomic region | No.Virus-encode miRNA targets | No. Host miRNA targets |
|---|---|---|
| 5'UTR | 2 | 2 |
| ORF1ab | 35 | 16 |
| S | 2 | 1 |
| ORF3a | 0 | 2 |
| E | 0 | 0 |
| M | 0 | 0 |
| ORF6 | 0 | 0 |
| ORF7a | 1 | 0 |
| ORF7b | 0 | 0 |
| ORF8 | 0 | 1 |
| N | 3 | 7 |
| ORF10 | 0 | 0 |
| 3UTR | 0 | 1 |

C

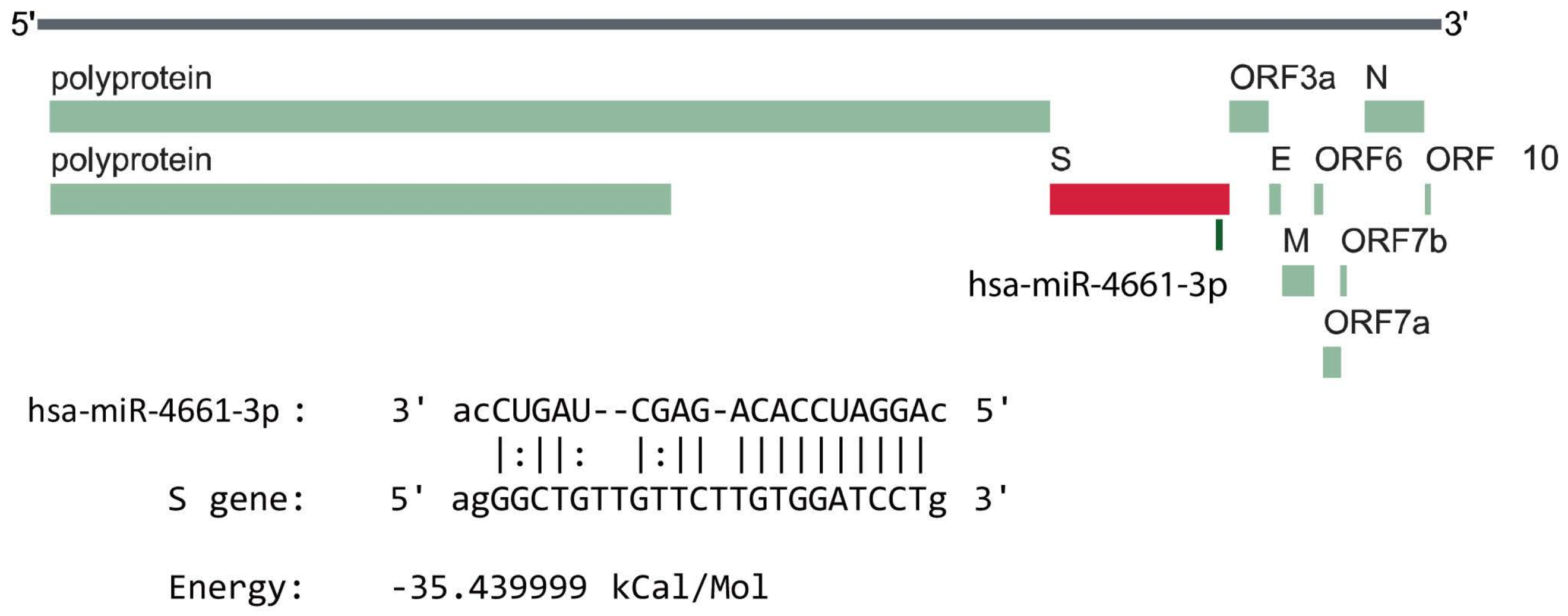

Figure 5. Virus genomic region targeted by virus and host-derived miRNAs. (A) The circos plot for interactions between virus miRNA and virus genomic region. (B) The distribution of viruses and host miRNA targets on the virus genome. (C). The target of hsa-miR-4661-3p at S gene.

Figure 6

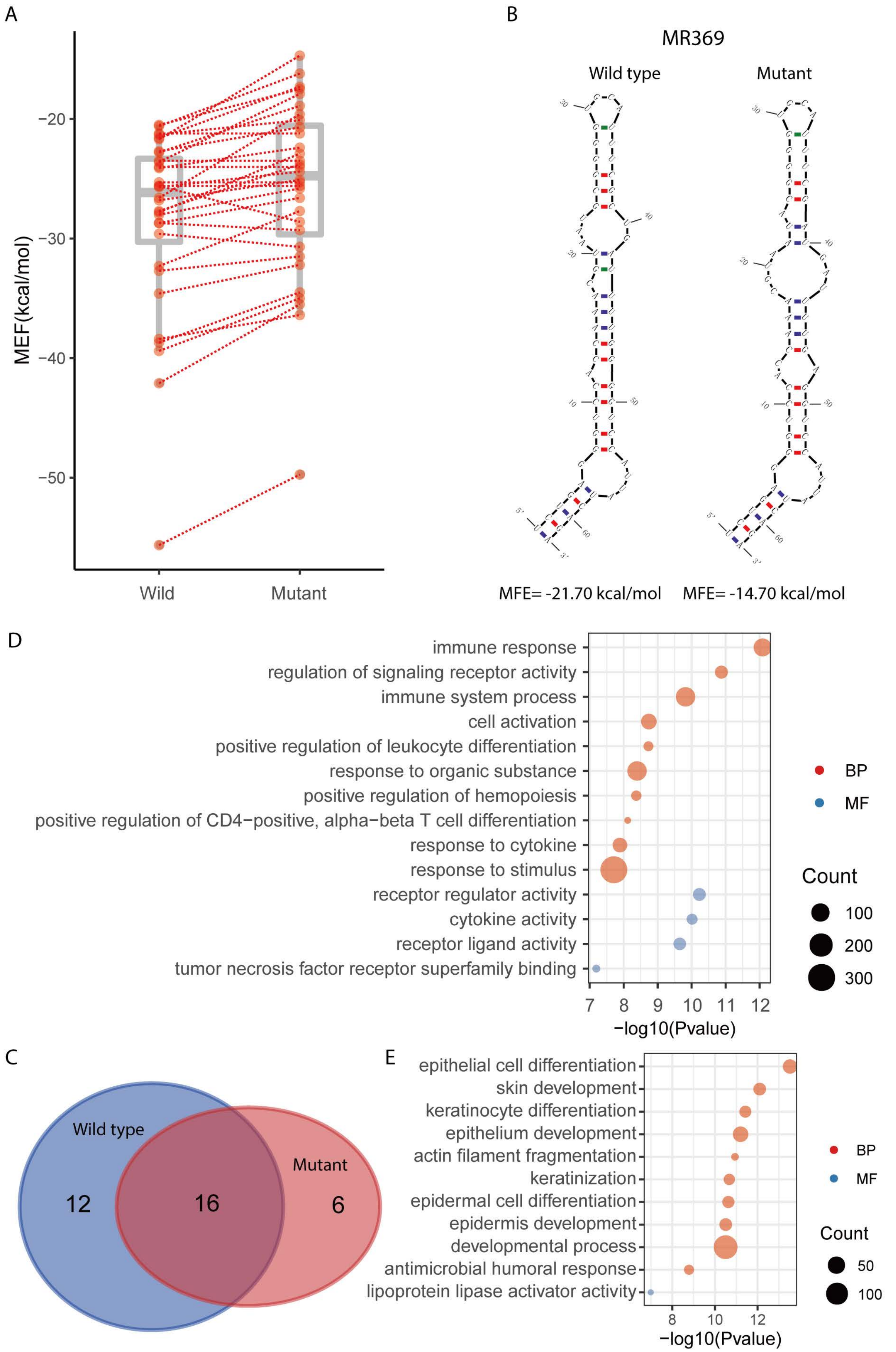

Figure 6. The effect of genome mutation on miRNA repositories and function. (A) The comparison between the MFE of hairpin folding of miRNAs derived from the wild type and mutated SARS-CoV-2 genome. (B) Diagram example of MR369 demonstrating the effect of mutations on folding energy. (C) The overlap of miRNAs that binding to the wild type and mutant genome. (D) The functional annotation of genes targeted by miRNAs specifically binding to the wild-type SARS-CoV-2 genome. (E) The functional annotation of genes targeted by miRNAs specifically binding to the mutant SARS-CoV-2

Figure 7

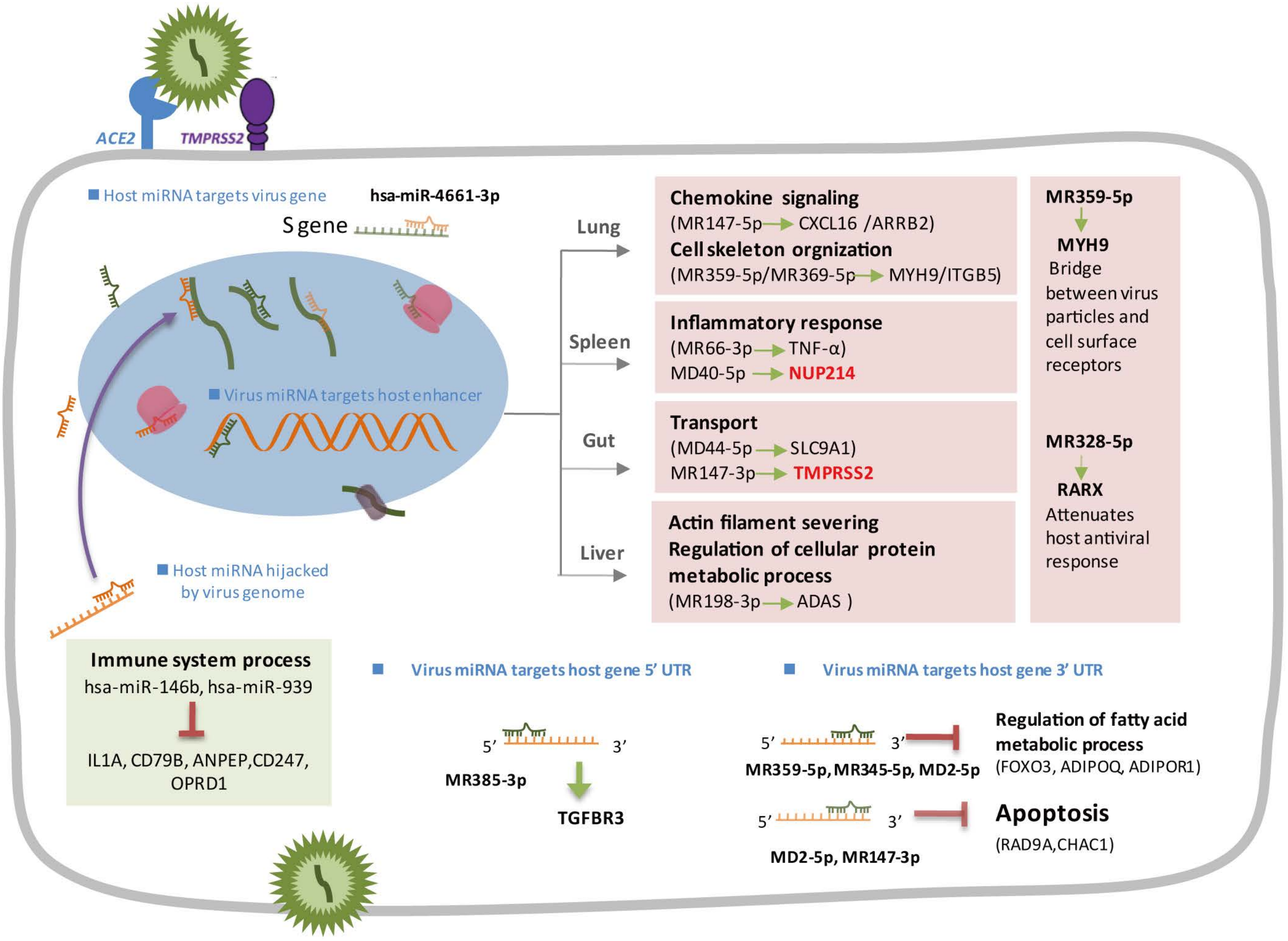

Figure 7. A graphic summary of the findings in this silico study.

A

Functions of virus-encoded miRNA

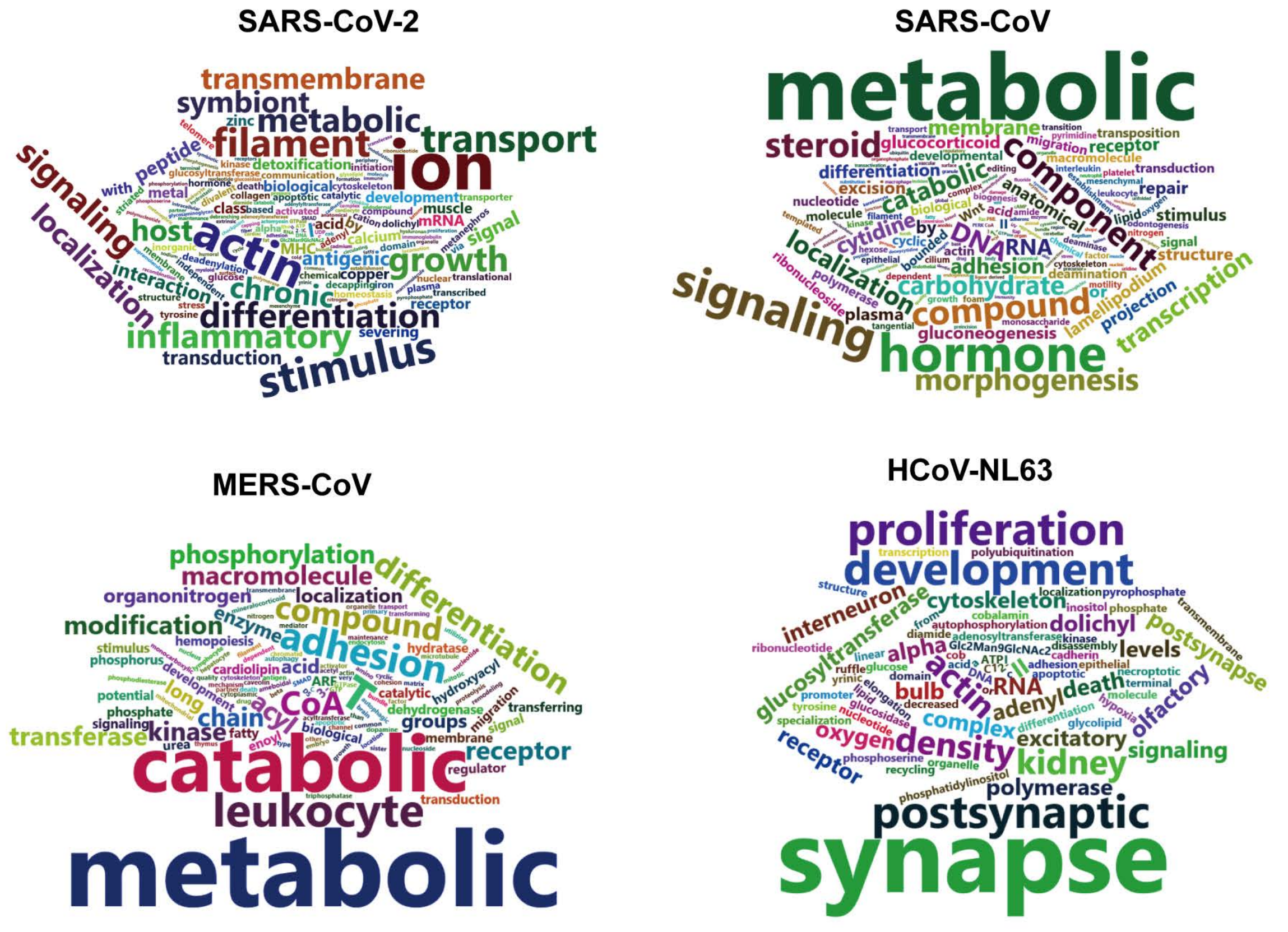

B

Functions of virus-binding miRNA

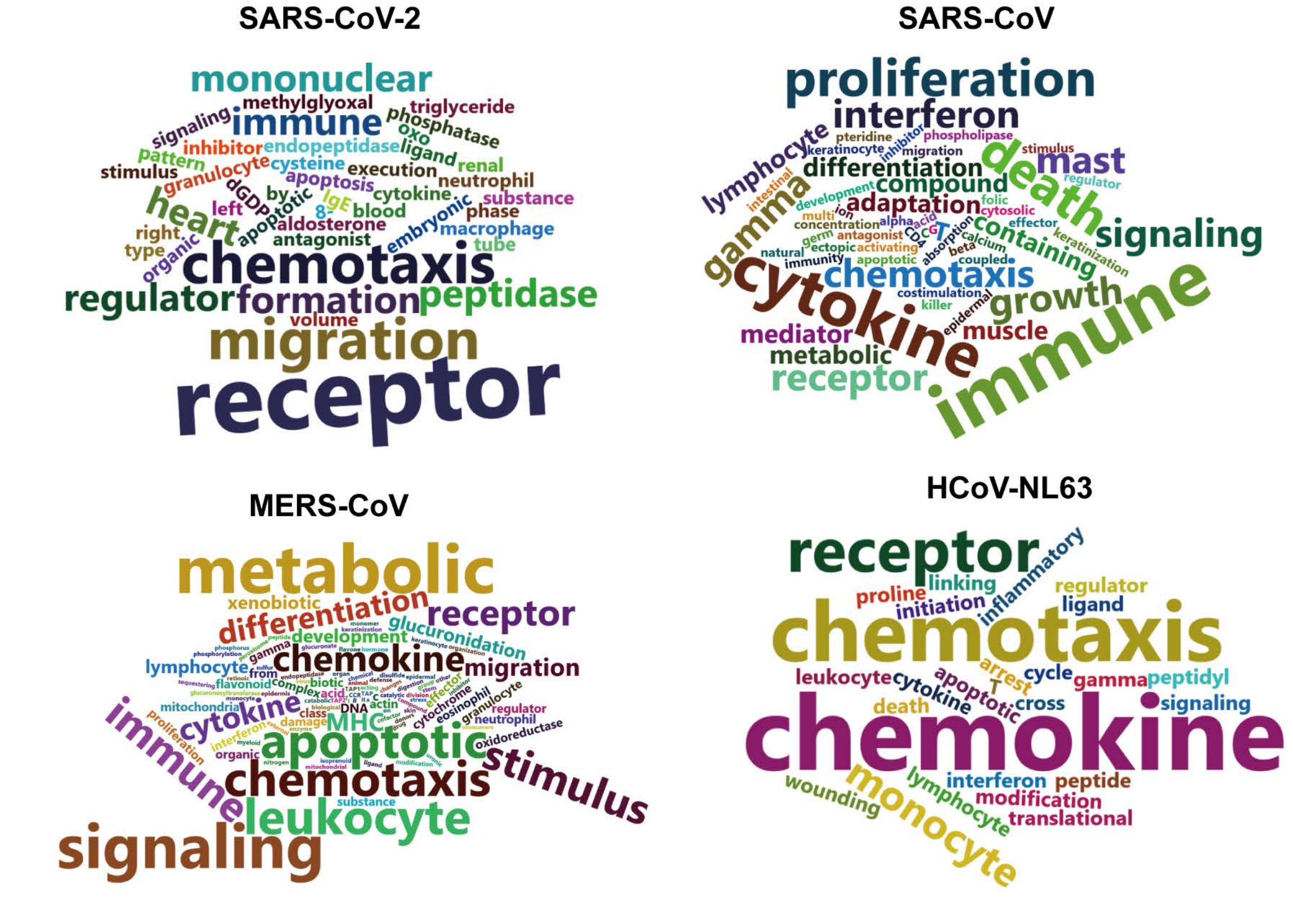

Figure S1. Functions of virus miRNA and host miRNA hijacked by the virus genome. (A) The wordcloud diagram of functions of gene targeted by virus miRNA on 3' UTR, 5' UTR and enhancers. The size of words is in proportion to the frequency of functions annotated to genes. (B) The wordcloud diagram of functions of genes targeted by host miRNA hijacked by virus genome.

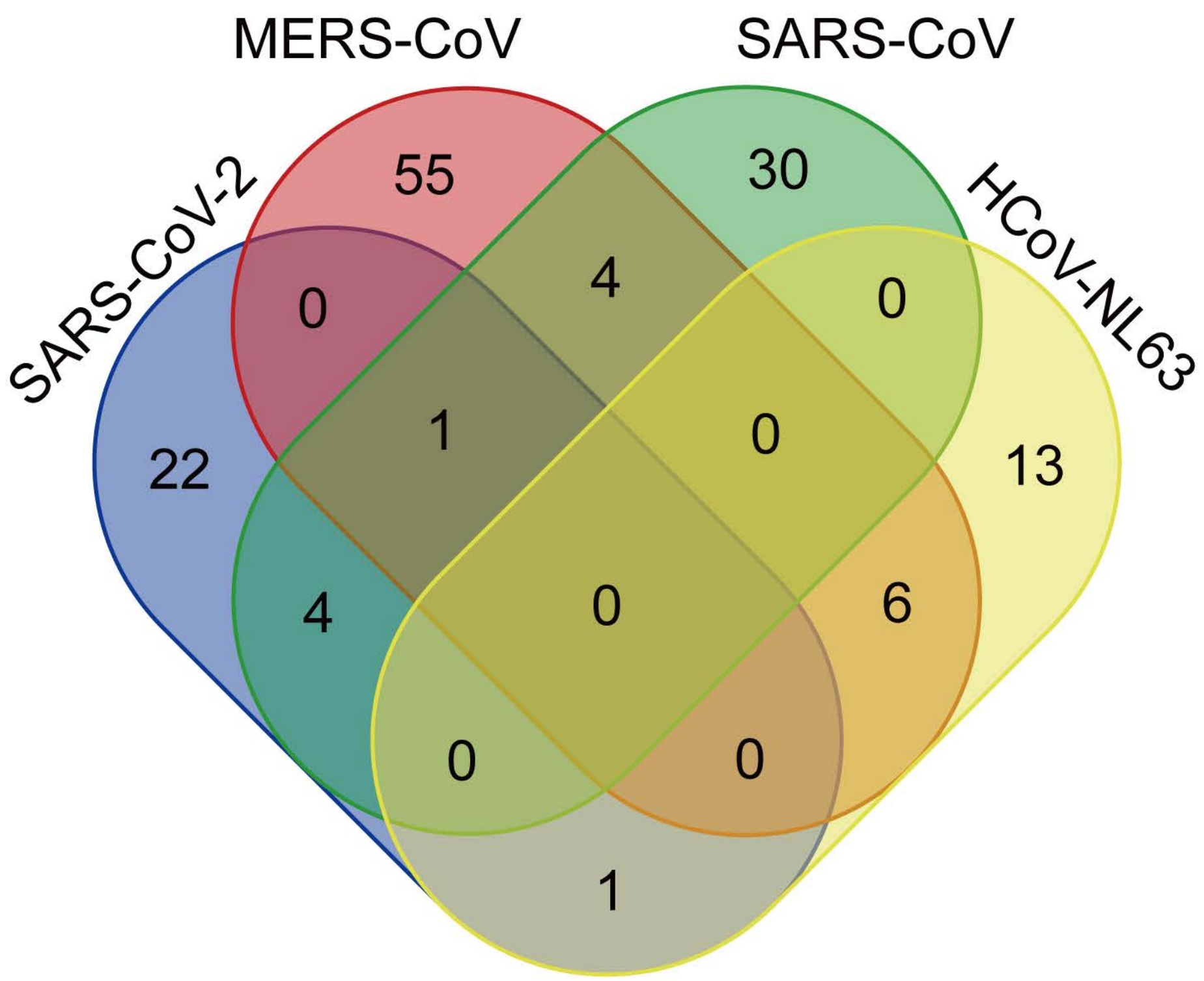

| Names | total | elements |
|---|---|---|
| MERS-CoV SARS-CoV SARS-CoV-2 | 1 | hsa-miR-4259 |
| SARS-CoV SARS-CoV-2 | 4 | hsa-miR-1181 hsa-miR-1307-3p hsa-miR-146b-3p hsa-miR-1229-5p |
| HCoV-NL63 SARS-CoV-2 | 1 | hsa-miR-1292-5p |
| MERS-CoV SARS-CoV | 4 | hsa-miR-4649-5p hsa-miR-6851-5p hsa-miR-345-3p hsa-miR-6885-5p |
| HCoV-NL63 MERS-CoV | 6 | hsa-miR-3085-5p hsa-miR-6831-5p hsa-miR-6089 hsa-miR-6812-5p hsa-miR-4695-5p hsa-miR-6782-5p |
| SARS-CoV-2 | 22 | hsa-miR-370-3p hsa-miR-6751-5p hsa-miR-623 hsa-miR-7161-3p hsa-miR-4786-3p hsa-miR-10398-3p hsa-miR-939-5p hsa-miR-193a-5p hsa-miR-6834-5p hsa-miR-324-3p hsa-miR-4738-3p hsa-miR-6882-3p hsa-miR-6827-5p hsa-miR-3154 hsa-miR-8075 hsa-miR-4526 hsa-miR-4758-5p hsa-miR-3191-3p hsa-miR-3648 hsa-miR-602 hsa-miR-6774-5p hsa-miR-4661-3p |
| SARS-CoV | 30 | hsa-miR-4515 hsa-miR-4739 hsa-miR-4687-3p hsa-miR-3137 hsa-miR-6090 hsa-miR-7158-5p hsa-miR-874-3p hsa-miR-328-5p hsa-miR-1193 hsa-miR-500a-5p hsa-miR-612 hsa-miR-935 hsa-miR-1204 hsa-miR-497-3p hsa-miR-622 hsa-miR-4418 hsa-miR-5572 hsa-miR-505-3p hsa-miR-1471 hsa-miR-4675 hsa-miR-4436a hsa-miR-214-3p hsa-miR-330-5p hsa-miR-1202 hsa-miR-6081 hsa-miR-6514-3p hsa-miR-12119 hsa-miR-658 hsa-miR-10401-5p hsa-miR-671-5p |
| MERS-CoV | 55 | hsa-miR-2682-5p hsa-miR-342-3p hsa-miR-637 hsa-miR-1538 hsa-miR-4646-5p hsa-miR-6840-3p hsa-miR-608 hsa-miR-4793-5p hsa-miR-1225-5p hsa-miR-6800-5p hsa-miR-339-3p hsa-miR-1224-5p hsa-miR-6781-5p hsa-miR-6741-5p hsa-miR-4787-5p hsa-miR-6787-5p hsa-miR-6855-5p hsa-miR-6824-5p hsa-miR-6796-5p hsa-miR-8060 hsa-miR-6865-5p hsa-miR-4640-5p hsa-miR-4691-3p hsa-miR-1228-5p hsa-miR-4743-5p hsa-miR-4717-5p hsa-miR-3194-5p hsa-miR-3689c hsa-miR-3138 hsa-miR-10396b-3p hsa-miR-4797-3p hsa-miR-5587-3p hsa-miR-6728-5p hsa-miR-11401 hsa-miR-3620-3p hsa-miR-6775-5p hsa-miR-5194 hsa-miR-6772-5p hsa-miR-12128 hsa-miR-5196-5p hsa-miR-4767 hsa-miR-134-5p hsa-miR-5001-5p hsa-miR-197-5p hsa-miR-3620-5p hsa-miR-4688 hsa-miR-4446-3p hsa-miR-3692-5p hsa-miR-3689b-3p hsa-miR-4298 hsa-miR-4498 hsa-miR-4632-5p hsa-miR-6799-5p hsa-miR-921 hsa-miR-3177-3p |
| HCoV-NL63 | 13 | hsa-miR-6750-5p hsa-miR-6780b-5p hsa-miR-6165 hsa-miR-4725-3p hsa-miR-6765-5p hsa-miR-506-3p hsa-miR-4518 hsa-miR-1913 hsa-miR-1183 hsa-miR-1295b-3p hsa-miR-5189-5p hsa-miR-4638-3p hsa-miR-6813-5p |

Figure S2. The comparison of miRNA targeted at the virus genome of SARS-CoV-2, SARS-CoV, MERS-CoV and HCoV-NL63.